\documentclass[a4paper]{article}

\usepackage[margin=1in]{geometry} 

\usepackage{amsmath}
\usepackage{amsthm}
\usepackage{amssymb}

\usepackage[utf8]{inputenc}
\usepackage{hyperref}
\hypersetup{
	unicode,
	pdfauthor={Jonathan Bright, Florence E. Enock, Pica Johansson, Helen Z. Margetts, Francesca Stevens},
	pdftitle={Understanding engagement with platform safety technology for reducing exposure to online harms},
	pdfsubject={Understanding engagement with platform safety technology for reducing exposure to online harms},
	pdfkeywords={Safety technology, User controls, User empowerment tools, Online Safety, Online harms, Survey research, Public attitudes },
	pdfproducer={LaTeX},
	pdfcreator={pdflatex}
}

\usepackage{apacite}

\usepackage{graphicx, color}

\title{Understanding engagement with platform safety technology for reducing exposure to online harms\footnote{This work was supported by the Ecosystem Leadership Award under the EPSRC Grant EPX03870X1 \& The Alan Turing Institute}}
\author{Jonathan Bright$^1$ \and Florence E. Enock$^1$ \and Pica Johansson$^1$ \and Helen Z. Margetts$^1,^2$ \and Francesca Stevens$^1$}

\date{
      $^1$Public Policy Programme, The Alan Turing Institute. \\ $^2$ Oxford Internet Institute, University of Oxford. \\ \texttt{Corresponding author: jbright@turing.ac.uk}\\[2ex]%
      \today
}

\begin{document}
	\maketitle
	
\begin{abstract}
User facing `platform safety technology' encompasses an array of tools offered by platforms to help people protect themselves from harm, for example allowing people to report content and unfollow or block other users. These tools are an increasingly important part of online safety: in the UK, legislation has made it a requirement for large platforms to offer them. However, little is known about user engagement with such tools. We present findings from a nationally representative survey of UK adults covering their awareness of and experiences with seven common safety technologies. We show that experience of online harms is widespread, with 67\% of people having seen what they perceived as harmful content online; 26\% of people have also had at least one piece of content removed by content moderation. Use of safety technologies is also high, with more than 80\% of people having used at least one. Awareness of specific tools is varied, with people more likely to be aware of `post-hoc' safety tools, such as reporting, than preventative measures. However, satisfaction with safety technologies is generally low. People who have previously seen online harms are more likely to use safety tools, implying a `learning the hard way' route to engagement. Those higher in digital literacy are also more likely to use some of these tools, raising concerns about the accessibility of these technologies to all users. Additionally, women are more likely to engage in particular types of online `safety work'. We discuss the implications of our results for those seeking a safer online environment. 
\newline

	\end{abstract}
	
	\section{Introduction}
	\label{sec:intro}

The potential for users of social media platforms to be exposed to harmful content remains a cause for concern. Numerous online harms, understood as harmful content or activity that is experienced online such as cyberstalking and hate speech (see HM Government, \citeyearNP{HMgov2019}) take place everyday. In the United States, Pew Research Centre \citeyear{Pew2021} found that 41\% of Americans have experienced online harassment, including physical threats, stalking and sexual harassment. In the United Kingdom, a recent report found that two thirds of the British public say they have witnessed harmful content online before, with the number rising to nine out of ten for young adults \cite{Turing2023_harmstracker}. Such high levels of exposure to online harms is problematic: these experiences have been linked to numerous negative psychological impacts including depression, anxiety, fear, low self-esteem and escalation of self-harm \cite{stevens2021cyber,susi2022viewing}. Furthermore, experiencing harms like online abuse and sexualised threats can result in victims withdrawing from social interaction, including with friends and family, employment and educational organisations \cite{pashang2018mental,Storry2022Impact}. 

Recently, social media platforms have faced growing legislative pressure to tackle harmful content. For example, in 2021, Australia introduced the Online Safety Act, which strengthens the country's laws around online safety, making providers more accountable and giving the eSafety Commissioner significant new powers to be able to protect Australians from harmful content online \cite{AusOnlineSafetyAct}. The EU's Digital Services Act aims to provide similar powers within the EU \cite{EUDigAct}. In the UK, the recently passed Online Safety Act puts forward a new regulatory framework to tackle online harms \cite{UKSafetyAct}. Importantly for the present article, a key requirement of the duty of care that platforms have to their users under this legislation is ensuring that all users have straightforward access to effective `safety technology', enabling user driven control of online safety. 

These `platform safety technologies', as we refer to them here, are tools that have been created to help people tailor their online experiences and protect themselves from harm online, especially on social media platforms. For example, mainstream platforms such as Facebook, Instagram and TikTok provide consumers with features that allow them to unfollow others \cite{bode2016pruning}, block other users from contacting them \cite{yang2017politics} or report content they feel is inappropriate to the platform \cite{crawford_what_2016}. 

While increased emphasis is now being placed on the importance of this type of safety technology in enhancing online safety, the utility of these features is dependent upon individuals knowing that they exist, choosing to use them, and understanding how to do so. However, there is currently little understanding of the extent of use of these platform safety technologies and the potential drivers of their use. We address this gap in this paper by describing levels of contemporary use of platform safety technology and explaining key determinants of this use. 

In the next section, we review existing work on the use of platform safety technology, and also review what is known about potential drivers of its uptake. We then present our method, a nationally representative survey of UK adults. We next present our results, showing both widespread awareness of and use of platform safety technologies, but also important demographic and experiential differences in terms of their uptake. We conclude by discussing the implications of these results for efforts to improve online safety. 

\section{Platform Safety Technology}
\label{sec:theory}
Since the emergence of social media platforms, the safety of their users has been a key focus for researchers, policymakers and practitioners; hence the extent of use of platform safety technology has attracted some research interest. Earlier work, which was typically focused on the use of safety technology for enhancing privacy, largely found usage levels to be low. One study \cite{acquisti2006imagined} exploring attitudes towards privacy on social media platforms found a disparity between users' wishes to protect their privacy compared with their actions: while many students expressed concern over their online privacy, less than 1\% modified the default profile visibility in their Facebook account. A subsequent survey found that the majority of social media users (76\%) were not aware of platforms’ privacy warnings \cite{ho_privacy_2009}. More recent research using qualitative methods to explore older teenagers' attitudes towards online privacy and safety also found that although over half of the participants encountered privacy settings on their social media accounts, few had actually reviewed or changed any \cite{agosto_dont_2017}. 

Recent research in the US examining user awareness, navigation, and the use of feed settings supports some of this earlier work. One study, which utilised an online survey along with in-person interviews, found that participants had difficulty locating feed settings, though they did wish to use them, with 94\% changing a minimum of one setting once they found where they were \cite{hsu2020awareness}. However, many users misunderstood the functions of such settings. For example, a quarter of participants believed that turning off advertising personalisation and removing advertising interests would lessen the quantity of adverts they were seeing on Facebook, rather than just changing their content. The study also showed that users of mainstream platforms such as Facebook, Twitter and YouTube were not aware of how much of their social media experience they could control by changing various settings. 

Recent work from Ofcom \citeyear{OfCom2022} reported moderate patterns of use of safety technology. In a nationally representative survey of UK adults, they showed that six in ten UK internet users who came across harmful content or behaviours online chose to use a safety feature in response to their experience. The most frequently used features were to unfollow, unfriend or block the person who had posted the content (20\%), as well as to make use of the report or flag button, or to mark it as junk (also 20\%). In other words, absolute usage of these features was low, showing that more needs to be done to engage users with these tools if they are to play a central part in enabling a safe experience online. While little is currently known about why people do and do not use safety technology, the complexity of reporting content in particular is something that has been highlighted in a variety of works as a reason why few people report content online \cite{voxpol2019p.69}. For example, reporting mechanisms may not be accessible enough on many platforms, users are often not given information about when and why content should be flagged, and there is often little transparency over what happens to flagged content \cite{crawford_what_2016}. Some work finds that users are more likely to flag potentially harmful content if they are provided with detailed guidelines on how to do so \cite{naab_flagging_2018}. Further work finds that enhancing transparency around how moderation systems work enhances users' trust in such systems \cite{molina2022ai}.  

In addition to trying to understand the extent to which safety features are used, previous studies have looked at the extent to which users are satisfied with the outcomes when they do make use of such features, something that has been addressed especially in the context of reporting of harmful content. Such studies have found that the overwhelming user experience on reporting content was that no action was taken - the content remained on social media platforms, with users also not receiving any response from the platform regarding their report. Hence, unsurprisingly, many users were ultimately dissatisfied with the reporting process \cite<e.g.,>{worsley2017victims}. A recent survey from Ofcom supports this, with less than half of users satisfied with the reporting process on social media as a whole \cite{OfCom2022}. 

Overall, however, the number of studies addressing questions of awareness, use and satisfaction with online safety technology is limited in scope. Findings of low usage levels might now be out of sync with the contemporary landscape which has seen an ever greater focus on online harms. Additionally, most work takes a narrow focus on only a small number of different types of safety technology and few studies employ large-scale, nationally representative samples to understand attitudes and behaviours which are generalisable to populations. This leads us to pose the first descriptive research question which motivates our work:
\begin{quote}
RQ1: What are the levels of awareness of, use of and satisfaction with platform safety technologies amongst users of social media platforms?
\end{quote}
In addition to this descriptive research question, we aim to explore potential drivers of the use of platform safety technology. Again, literature in this area is relatively limited in scope. However, some work exists that has built on more general work on theories of safety behaviour, which allows us to develop some hypotheses about drivers of use. 

A first area of research has drawn from Protection Motivation Theory \cite{rogers1975protection}. This theory was originally conceived to explain how individuals behave in a self-protective way in response to a perceived health threat and it has since been applied to users protecting themselves online. Research has demonstrated that improving users' perceptions of their personal responsibility to protect themselves is needed first and foremost in order for effective use of online safety interventions, as well as the condition that the user's understanding should correspond to the intervention strategy \cite{shillair_online_2015}. In other words, concerns about online harms are one motivator for users to engage with platform safety technology. This leads to our first hypothesis:
\begin{quote}
H1: Individuals who are more concerned about online harms will be more likely to use use platform safety technology. 
\end{quote}
Of course, one way of becoming concerned about online harms is through direct experiences of harms. This is supported by research which found that past experiences of harms (in this case privacy breaches) are associated with higher levels of protective behaviours, which implies a `learning the hard way' approach exists when it comes to engaging with platform safety technology \cite{buchi2017caring}. This strand of work leads to our second hypothesis:
\begin{quote}
H2: Individuals with prior experience of online harms will be more likely to use platform safety technology. 
\end{quote}
In addition to experience with harms, skill and ability with the use of technology is also likely to play a role. One lens through which this has been explored is through the concept of `digital pruning', which explains the method of `sifting through and unfollowing content that triggers undesirable affect and negative state of mind', with users describing it as `an act of self-care, requiring sustained reflection and evolved self-knowledge' \cite[p.~13]{hockin-boyers_digital_2021}. The concept resulted from qualitative research with female participants who were weightlifting during their recovery from an eating disorder on how they navigate social media. Digital pruning was described by participants as a skill, and supports more general work on the idea that those making the most use out of platform safety technology might be the most digitally skilled. \citeA{eastin2000internet} created an internet self-efficacy scale, consisting of eight items including users' confidence with understanding terms relating to Internet software and learning advanced skills within a specific Internet program. The survey, completed by American college students, found that previous internet experience, outcome expectations and internet use were all significantly and positively associated with individuals' internet self-efficacy beliefs, which we might expect to lead more generally to the use of platform safety technology. This was supported in a study of reporting behaviours, which showed that that self-efficacy and expectations of success were key drivers in the use of reporting functions \cite{wong_standing_2021}. These lines of thought lead us to our next hypothesis:
\begin{quote}
H3: People with greater digital literacy will be more likely to use platform safety technology.
\end{quote}
There is also research providing some suggestion that there may be gendered effects on the likelihood of using platform safety technology. Part of this may be driven by differing exposure to online harms. Although this is addressed already in hypothesis 1, there is evidence to suggest that the type of online harms women are exposed to are qualitatively different to those that men are exposed to (not enough is known about how this also impacts people who identify as non-binary). For example, women may be more directly exposed to threats of cyberaggression \cite{levrant2001cyberaggression}, or direct sexualised threats of `intimate intrusion' \cite{gillett2023not} that might be particularly likely to provoke the use of platform safety technology, compared to other types of online harms (for example, exposure to scams or online phishing attacks). Whether this leads to differing use of platform safety technology is an open question: some work has shown gender differences in this context \cite{park2013offline}, whilst other work has not found a clear effect \cite{wilhelm_gendered_2019}. This leads us to our fourth hypothesis:
\begin{quote}
H4: Women will be more likely to use platform safety technology than men.
\end{quote}
Finally, there is some research supporting the idea that political orientation may play a role in engagement with safety technology. One way of looking at the use of platform safety technology is in terms of its potential impact on free speech, an area where there are clear partisan differences. In particular, in the context of misinformation reporting, a variety of work has suggested that left-wing people are more likely to be in favour of actions that restrict misinformation, with right-wing individuals more likely to perceive this as an assault on free speech (Pew Research Centre, \citeyearNP{Pew2023}). Empirical work on reporting behaviours has thus far not found any influence of partisanship \cite{wilhelm_gendered_2019, riedl2022antecedents}, however, to our knowledge the relationship has not been tested for other types of platform safety technology. This leads us to our final hypothesis: \begin{quote}
H5: Left-wing users will be more likely to use platform safety technology.
\end{quote}
In addition to our main hypotheses, we also include a number of control variables in our analyses that are likely to be correlated with some of the variables in our main hypotheses. Frequency of social media use has an obvious potential effect, with those using the technology less being perhaps less likely to encounter harms. Level of education is a further factor which may be correlated with level of digital literacy. Finally, age is also a factor that has been mentioned frequently in the literature, with older users perhaps using social media less frequently and also less digitally literate than younger people \cite{dodel_inequality_2018, park2013offline}. 

Despite the existing work discussed in this section, considerable gaps in the research landscape remain. Much of the literature referenced here is out of date or focuses only on privacy and cybersecurity. Engagement with newer platform safety technologies, such as controlling the order of one's feed, have received only little attention. Furthermore, explanatory models of why people use these critical technologies are only loosely developed. Our work seeks to fill these important gaps.  

\section{Data and Methods}
\label{sec:dataandmethods}

In this section, we describe our approach to data collection and the creation of our measures. Our main research instrument was a nationally representative survey of UK adults. Respondents were recruited through the platform Qualtrics\footnote{www.qualtrics.com}, which was also used to create and administer the survey, and data was collected during January 2023. The entire survey was designed to take each respondent approximately twenty minutes to complete, and participants were remunerated at a standard rate for their time. A total of 1,067 participants who completed the survey passed standard checks for data quality and were included in the final sample. The sample was designed to be nationally representative of the population of the United Kingdom across demographic variables of age, gender and ethnicity. The demographic breakdowns used as standard by Qualtrics are based on 2020 data from Eurostat. A full breakdown of the demographics of the participants can be found in Appendix \ref{sec:demographics}. The survey was approved by the Ethics Committee at The Alan Turing Institute, UK (approval number C2105-074). Informed consent was obtained at the start of the survey according to approved ethical procedures. 

\begin{table}
    \centering
    \begin{tabular}{p{0.35\linewidth} p{0.6\linewidth}}
        Preventative actions (exposure) & \\
        \hline
        Algorithmic feed modification& The ability to personalise what you see on your feed. You can do this using particular keywords, for example to indicate that you would like to see more or less of a certain type of content.\\
        \\
        Algorithmic feed deactivation& The ability to completely deactivate any personalisation of feed content, switching to purely a chronological order\\
        \\
        Preventative actions (posting) & \\
        \hline
        Limit responses&Limiting who can respond to your posts means you choose who can interact with your posts, whether it is anybody (public), only people that follow you (private), or personalised for specific users.\\
        \\
        Hide responses& The ability to hide reactions (such as `likes') and replies to content on social media, both for your own posts and for other people's posts that you may view. \\
        \\
        Post-hoc measures & \\
        \hline
        Unfollow& Unfollowing another user is when you unsubscribe from receiving their content, meaning it will no longer appear on your feed. \\
        \\
        Block&Blocking a user is when you prevent another user from being able to follow you, view your content, or contact you on social media.\\
        \\
        Report& The ability to report content which violates platform guidelines, including offensive and hateful material. This may involve reporting another user, group, a particular piece of content, or a private message.\\ 
    \end{tabular}
    \caption{Platform safety technologies, with descriptions as presented in the survey}
    \label{tab:safety_technologies}
\end{table}

Our key dependent variables concern people's awareness of and experience with platform safety technologies. We sought to create an exhaustive list of common safety technologies that users were likely to have encountered on major social media platforms, and to ask participants about each of them. The precise technologies we focus on are presented in Table \ref{tab:safety_technologies}, along with the descriptions that we presented to the survey participants. These seven safety technologies fall into three broad categories: preventative technologies that might limit exposure to potentially undesirable content (algorithmic feed modification and algorithmic feed deactivation); preventative technologies that limit undesirable consequences from posting content (limiting responses and hiding engagements); and post-hoc technologies that could be taken in response to having seen undesirable content or experienced undesirable interactions, with the aim of preventing re-occurrence (unfollowing, blocking and reporting).

It is worth noting that some of these features (especially unfollowing and the ability to modify or deactivate an algorithmic feed) are not only safety focused, and might be used more generally to personalise a user's online experience. It is also worth noting the diverse types of harm that these features could protect from: for example, hiding engagements is aimed at protecting mental health and reducing some of the social pressure of operating on a social platform, whereas blocking and reporting are aimed at directly reducing experiences with harmful content created by other users.  

For each of the seven features, we asked participants whether they had heard of them, and how much they had previously used the feature; if they had, we asked whether the outcome was as they had hoped. In addition to the main questions about use of safety features, we furthermore asked about experiences with content removal on social media. Participants were asked whether they had ever had content removed on social media before; if they had had content removed before, whether they appealed the removal; and if they previously appealed, whether the outcome was as they had hoped.   

We also collected data for the independent variables in our study. Participants indicated how concerned they feel about harmful online content using an unmarked sliding scale (scored from 0-100) from `Not at all concerned' to `Extremely concerned', allowing us to address hypothesis 1. 

We asked participants the extent to which they had been exposed to content which they consider to be harmful on social media in the past along two scales: how much they had witnessed content which they consider to be harmful online, and how much they had directly received content that they consider to be harmful online (‘Many times’, ‘Occasionally, from time to time’, ‘Very rarely, once or twice’, ‘Never’). This allowed us to capture a distinction between, for example, witnessing hate speech directed at a celebrity, or directly receiving hate speech online. Both of these variables address hypothesis 2.  

We measured digital literacy (hypothesis 3) by asking participants to indicate how confident they feel using computers, smartphones, or other devices to do eight different activities online, including sending and receiving emails, organising content across multiple devices, and purchasing goods online. Participants responded to the eight items using unnumbered sliding scales (scored from 0-100), with guides for `Not at all confident', `Not very confident', `Fairly confident' and `Very confident' equally spaced along the sliders. These items showed good internal reliability ($\alpha$ = 0.85) and we combined these using the mean to create single digital literacy scores for each participant.

We collected demographic information about gender and political orientation, allowing us to test hypothesis 4 and hypothesis 5. For gender, participants were asked to select the option that best described them from a list of standard predefined categories, as well as being given an option to enter their own gender if they felt none of the categories applied. For political orientation, participants used an unmarked sliding scale (scored from 0-100) to indicate their political ideology from `extreme left' (0) to `extreme right' (100), with `centre' being a mid-point score of 50.

We collected further information to include as control variables. Participants were asked how often they used social media, with the resulting variable transformed into a dichotomous one indicating whether participants spent more than one hour per day on social media. Education level was presented as a standard list of predefined categories, that we transformed into a dichotomous variable indicating whether people had gone through university level education or not. Age could be entered as any number (though to emphasise again the survey was not open to those under 18). 

Gender, political orientation and education level all came with a `prefer not to say' option. For those respondents who selected `prefer not to say', their responses were set to the mean or modal category of the given variable as appropriate.   

A link to the full list of questions asked in the survey is available in Appendix \ref{sec:questions}, as well as key descriptive statistics (\ref{tab:iv_desc}). Although it was not the primary focus of the study, one finding worth highlighting from these descriptive statistics was the relatively high levels of perceived exposure to online harms: approximately 67\% of people said they had seen content online that they perceived to be harmful, with 31\% having directly received it. This supports statistics highlighted in the introduction above about the relatively widespread nature of interactions with online harms. 

\section{Results}
\label{sec:results}

\begin{figure}
    \centering
    \includegraphics[width=1.0\linewidth]{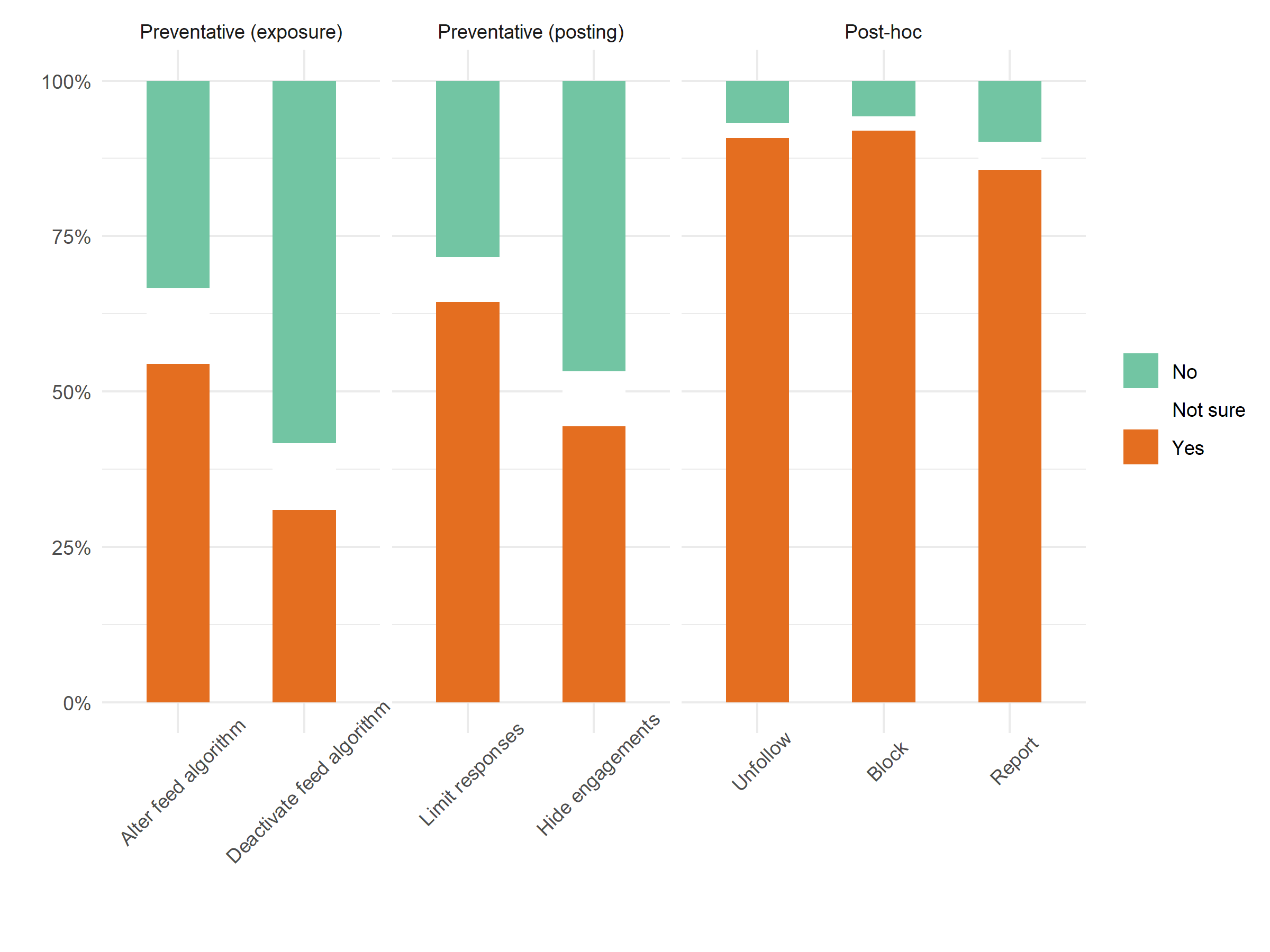}
    \caption{Awareness of platform safety technology}
    \label{fig:awareness-controls}
\end{figure}

The initial descriptive focus of this study was to explore levels of awareness, use and satisfaction with online safety technologies amongst users of social media platforms. Figure \ref{fig:awareness-controls} reports our results for safety technology awareness (the data behind the figure is available in Appendix \ref{sec:app_desc_stats}, Table \ref{tab:fig1_data}). 

Awareness was generally lower for the technologies focused on preventing exposure to harms before they occur, with only 40\% of people aware of options to deactivate feed algorithms, and 53\% aware of the option to hide engagement on content they create. Awareness was highest for post-hoc measures, with 86\% of people aware of reporting mechanisms and more than 90\% aware of options to block and unfollow. Overall, we find that awareness of online safety technologies is high in absolute terms, updating some of the less recent work we have mentioned above that finds awareness levels to be quite low. However, awareness is clearly higher for post-hoc platform safety technologies than it is for preventative safety technologies.   

\begin{figure}
    \centering
    \includegraphics[width=1.0\linewidth]{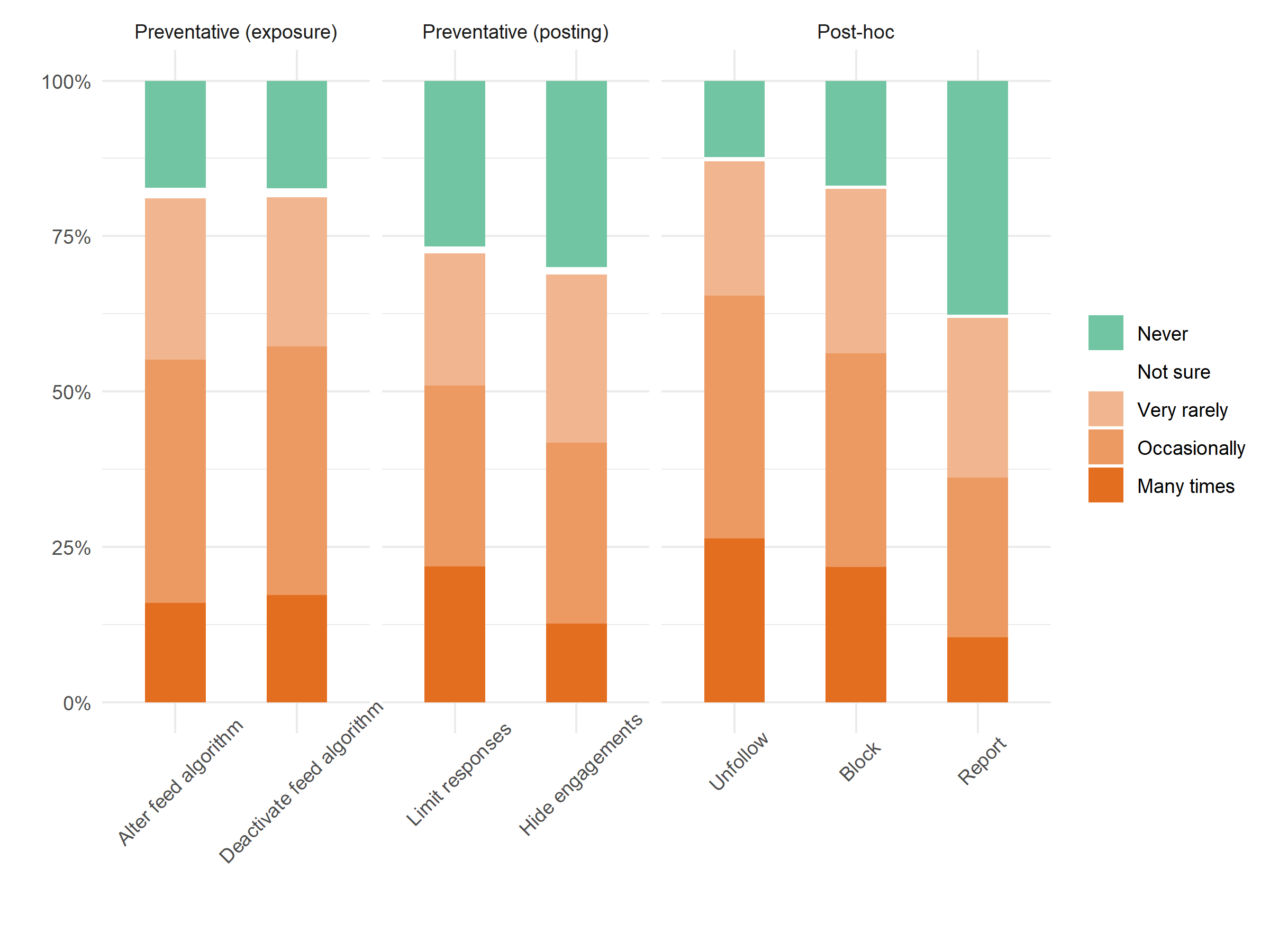}
    \caption{Use of platform safety technology}
    \label{fig:use-controls}
\end{figure}
 
Respondents were then asked to what extent they had (if at all) used each of the tools, with the results presented in Figure \ref{fig:use-controls} (again the full table is available in Appendix \ref{sec:app_desc_stats}). Note that the data in this figure is limited only to those people who had already said that they were aware of the technology. 

In usage terms, we see much less of a distinction between the preventative technologies and the post-hoc measures. Unfollow (more than 85\%) and block (over 80\%) were the most commonly used technologies, but although reporting was something people were widely aware of (as we show above in Figure \ref{fig:awareness-controls}), usage levels were lower, with only around 60\% of those aware of the technology ever having made use of it (though this is still a substantial amount in absolute terms). For the preventative technologies, in each case more than 70\% of those aware of them had used them. In other words, once people become aware of these technologies, they also tend to use them at relatively high rates. Again, this provides a counter point to results above that had suggested relatively low rates of usage of platform safety technology, and suggests that user behaviour is shifting. In total (including all the people in our sample, not just those who were aware of the technologies) more than 80\% of those surveyed had used at least one platform safety technology. 

Respondents were also asked about the extent to which they were satisfied with the outcomes of platform safety technology if they did make use of it. The extent of satisfaction varied for post-hoc technologies, with blocking and unfollowing attracting relatively high levels of satisfaction (more than 60\% of people said they were always satisfied with the outcome), compared to reporting where just 18\% of people said they were always satisfied with the outcome. This supports findings above showing that those using reporting technology were rarely satisfied with it. Preventative technologies also varied in terms of satisfaction, with 50\% of people saying they were happy with the results of limiting engagement, and 30-35\% of people saying they were happy with the results of altering the feed algorithm, deactivating the feed algorithm and hiding engagements on their content. Overall it is striking that while these safety technologies are widely used, satisfaction with their results is relatively low.  

We also asked about personal experiences with enforcement actions that platforms themselves were taking. This provides a different perspective on the prevalence of platform safety technology: if usage of reporting mechanisms is high, then deletion of content should also be high (though, of course, not all deleted content will have been deleted because of a user report - platforms also have their own internal content moderation mechanisms). Approximately 26\% of respondents said that they had had a piece of content deleted by a social media platform at least once, with around 16\% saying it had happened more than once. This indicates the widespread nature of content moderation. Of those who had had content removed, approximately 48\% launched a subsequent appeal process, and 30\% of these people received what they regarded as a satisfactory response. This in turn shows how contestable these decisions are, and how often platforms have to overturn enforcement decisions. It is interesting to compare this with the experiences of those reporting content, who are in general less satisfied than those who appeal content moderation decisions. However, it is worth noting that our sample of people who had been through an appeals process is relatively small, just 278 people. 

We will now move on to our main analytical task, which is explaining variation in use of platform safety technology. In order to approach this analysis we make use of a set of seven logistic regressions. For each regression, the dependent variable is a dichotomous indicator for whether respondents to the survey indicated that they had ever used one of the seven platform safety technologies under study (with respondents who said `Not sure' included in the `No' category). Each of the regressions looks at only the subset of respondents who were aware of the safety technology, hence the model seeks to explain use given that people have already heard of the technology (meaning that sample sizes vary from just over 300 to just under 1,000, depending on the technology in question). 

The models used to report these results were run through standard diagnostic tests for multicollinearity and influential values. The checks for influential values highlighted some points with high leverage, as measured by Cook's distance. Further versions of all models were produced with these high leverage values truncated, producing largely similar results (with only one difference, which is highlighted in the text below). Overall the diagnostics produced no reason to doubt the results. Full regression tables with accompanying Cox-Snell $R^2$ values and sample sizes are reported in Appendix \ref{sec:app_reg_results}, Tables \ref{tab:fig3_prevent} and \ref{tab:fig3_posthoc}. It is worth highlighting that the Cox-Snell values are low (less than 0.2 in all cases). 

\begin{figure}
    \centering
    \includegraphics[width=1.0\linewidth]{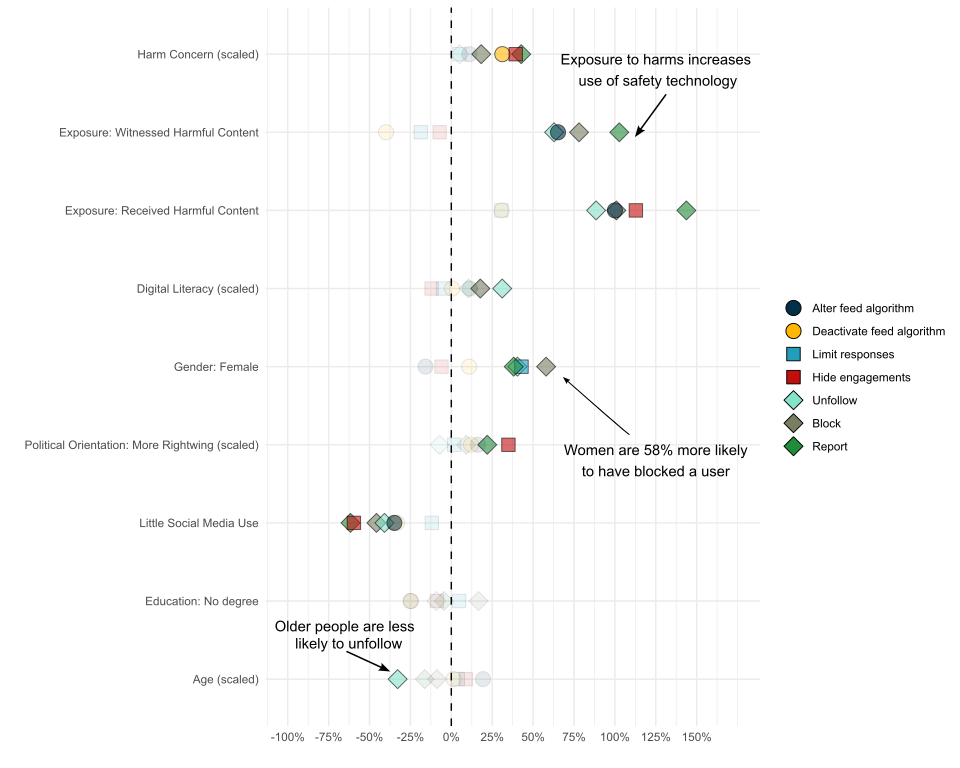}
    \caption{Explaining usage of platform safety technology}
    \label{fig:logistic_regression}
\end{figure}

Figure \ref{fig:logistic_regression} displays the results from these seven regressions. Each point on the graphic represents a coefficient from one of the regressions, with the transparent points statistically insignificant. Coefficients are exponentiated, meaning that they can be interpreted in terms of percentage increase or decrease in being more likely to have used a given safety technology. So, for example, women are about 58\% more likely to have blocked someone. 

A number of results stand out from the regressions. First, concerns about harm motivate use of technologies, supporting hypothesis 1. Those more concerned about harm were more likely to use preventative measures (deactivating the feed algorithm, hiding engagements) and post-hoc measures (blocking, reporting). Effect sizes were mixed: an increase of one standard deviation on the harm scale resulted in increases in likelihood of using a technology ranging from 18\% (blocking) to 43\% (reporting) depending on the technology.

However, even accounting for concerns, prior exposure to online harms also makes a considerable difference, supporting hypothesis 2. There were clear effects here for all of the post-hoc technologies: people who had either witnessed or been directly exposed to online harms were more likely to have unfollowed, blocked and reported (with effects in the range 78\% to 144\% depending on the technology). There were also results for some preventative technologies: altering the feed algorithm was more likely for those who witnessed harmful content (65\%) and for those who directly received it (100\%), while hiding engagements was more likely for those who had directly received harmful content (113\%). This shows good support for a `learning the hard way' effect in the use of safety technology.  

Digital literacy makes no difference for the preventative technologies, but does seem to have an impact on post-hoc technologies, where a one standard deviation increase in literacy is associated with an 18\% increase in the likelihood of blocking and a 31\% increase in the likelihood of unfollowing. It is perhaps concerning that there are literacy differences in use of safety technologies, though the effect is at least limited only to post-hoc technologies. So overall hypothesis 3 is only partially supported. 

There is more support for hypothesis 4, especially so for post-hoc technologies, with women more likely to report (38\%), unfollow (40\%) and block (58\%), though the result for unfollowing became insignificant in the model without outliers. Women were also more likely to limit engagements on their content (43\%). This result is important because, even accounting for both concerns about harms and exposure to harms, there are diverging gendered experiences, with women more likely to engage in some types of online `safety work', as recognised by the literature above. 

There are mixed results for political orientation. Users who are more right-wing are more likely to report content (22\% per one standard deviation increase on the scale) and hide engagements (35\% per one standard deviation), directly contradicting hypothesis 5. However the results are not present for the other technologies and so overall there is only weak evidence of a partisanship effect, and no support for hypothesis 5 (which predicted that left-wing users would make more use of platform safety technology).

Finally it is worth briefly commenting on the control variables. Infrequent users of social media were less likely to report, hide, block, unfollow and alter the feed algorithm, as we expected. Older users were less likely to unfollow but other than that age had no impact. There were no significant results for education level. 

\section{Discussion}
Engagement with online safety technologies is an increasingly important part of enabling a safer online environment for social media users \cite{UKSafetyAct}. However, the extent to which people feel able to protect themselves from experiencing online harms, and the extent to which people may find these tools more or less accessible, has so far been unclear. Using a large, nationally representative survey of UK adults, we examined people's awareness of, experiences with and attitudes towards seven common safety technologies currently found on social media platforms. We included technologies that aim to prevent exposure to potentially undesirable content (algorithmic feed modification and algorithmic feed deactivation); technologies that aim to prevent undesirable consequences from posting content (limiting responses and hiding engagements); and post-hoc technologies that may be used in response to having seen undesirable content (unfollowing, blocking and reporting). For each safety technology, we analysed overall proportions of awareness and experience and modelled key demographic and attitudinal predictors of engagement. 

When we examined the extent to which people said they had been exposed to harmful content online before, we found that our investigation into self-protection against such harms was necessary - 67\% reported having seen harmful content before, and 31\% reported having been directly sent harmful content, while overall concerns about online harms were correspondingly high, corroborating other work in this area \cite{Turing2023_harmstracker}. It is important to gain an up-to-date understanding of the extent to which people are exposed to online harms, particularly as some work suggests a dramatic increase in the prevalence of some type of online harms in recent years and since the Covid-19 pandemic \cite{Brandwatch2021}. 

Overall awareness of the safety technologies varied across the seven tools. While a large majority (90\%) had heard of options to block and unfollow other users, only around 40\% had heard of being able to deactivate their feed algorithms. This shows that options like this, designed to prevent harm exposure rather than respond to it, should be made more salient to social media users. Some usage levels were high - of the people who had heard of being able to unfollow other users, over 85\% had done so, while for blocking this was over 80\%. However, only 60\% of those who had heard of being able to report content had done so. We find some usage levels to be higher than prior work reports. For example, a recent report from Ofcom \citeyear{OfCom2022} found that just 20\% of users had flagged potentially harmful content which they had encountered online, compared to our 60\%. This disparity shows the importance of continuous up-to-date work as people's behaviours may be quickly changing with increased focus on and public conversation around online safety. 

Despite absolute use of safety technologies being high, people indicated generally low satisfaction with outcomes of use. This is particularly true when the outcome relies on a platform taking action (such as for reporting, where only 18\% say they were satisfied with the outcome) compared to a result happening automatically (such as for blocking another user, where more than 60\% say they were satisfied with the outcome). These findings are in line with previous work examining user engagement with online reporting, often finding that a lack of transparency around what happens to flagged content, along with low rates of platform action in response to user reports, means that users are frequently disappointed in reporting processes on social media platforms \cite{crawford_what_2016, OfCom2022, worsley2017victims, naab_flagging_2018, molina2022ai}. Our results show it is imperative for platforms to ensure that the safety features they offer are transparent and effective in order to maintain user engagement with playing a part to protect both themselves and their communities online.

In terms of demographic and attitudinal predictors of engagement with safety tools, it is interesting that prior exposure to online harms significantly predicts the use of several safety technologies, including all of the post-hoc technologies (unfollow, block, report) and also altering feed algorithm. This suggests that people often learn the hard way when it comes to engaging with safety technology, using these tools as a response to experience of harm rather than to preempt and prevent harm experiences from happening, supporting work elsewhere which shows experiences of harms lead to higher levels of protective behaviours \cite{buchi2017caring}. This result highlights the importance of encouraging preventative safety behaviours amongst social media users such that they can reduce their risk of being exposed to harmful content before it is too late, rather than having to take reactive measures against it. Media literacy courses and similar initiatives could educate people on using preventative safety tools, and platforms could make these settings more salient in the user environment. Digital literacy scores made some difference to engagement with safety tools, with those higher in digital literacy more likely to use post-hoc measures such as blocking and unfollowing, though with no effect for preventative measures, so only partially supporting prior work showing that digital skills enable self-protective online behaviours \cite{hockin-boyers_digital_2021, eastin2000internet, wong_standing_2021}. While it is somewhat positive that digital literacy does not play a large role in the use of safety technologies, indicating that these technologies are equally accessible for those high and low in digital skills, it is important to note that digital literacy scores were generally high in our sample. The items included in the scale we used are relatively simple (for example, being able to send an email) and our sample was drawn from people able and willing to take part in an online survey. It is possible that these effects would be stronger with more variability in digital literacy amongst respondents. 

It is interesting to note that female participants were more likely to engage with some of the safety tools than male participants, particularly for post-hoc measures (unfollowing, blocking and reporting) and also for limiting engagements on their posts. While there is conflicting evidence regarding gender differences in absolute exposure to online harms, with some work finding no differences in overall exposure \cite<e.g.,>{Turing2023_harmstracker}, it is likely that harms affecting women online are qualitatively different to those affecting men, with women more at risk of `contact' harms such as sexual harassment and cyberflashing \cite{levrant2001cyberaggression, gillett2023not}. It may be that the specific kinds of harms women experience encourage them to do more safety work in this area \cite{park2013offline}, thus leading to higher engagement with some safety technologies, particularly ones which focus on controlling how other users can interact with their profile and content. It may also be that fear surrounding the experience of particular online harms leads women to take at least some preventative measures to a greater degree than men. Future work will benefit from further understanding the specific ways in which women are affected by online harms along with how this alters their online behaviours. 

Our work makes several important contributions to the literature surrounding online safety behaviours amongst social media users. However, we note some limitations in our design. Firstly, for the models predicting engagement with each of the seven safety tools, we included only respondents that had heard of the safety feature in question, so sample sizes for our seven models varied substantially, from just 330 for deactivating feed algorithm, to 981 for blocking other users. Further, we did not ask which particular online harms people have been exposed to, which may also play an important role in the kinds of safety behaviours users engage with. Finally, our sample included only adults, and future work will benefit from understanding engagement with online safety technologies in children and teenagers, who are in many cases particularly at risk of suffering the adverse effects of online harm exposure \cite{livingstone2014risk, mchugh2018social}. 

Overall, we provide up-to-date evidence on general exposure to online harms and experiences with safety technologies in a large, nationally representative sample of UK adults. In particular, we highlight the importance of encouraging awareness and use of preventative safety behaviours as well as reactive ones, and we reiterate that platforms must make more effort to make safety technologies transparent, accessible, and effective if users are to engage. Understanding the social and psychological mechanisms underlying user intervention against online harms is a crucial step towards ensuring a safer online environment. By deepening our understanding in this area, our work also provides a strong foundation for further research testing interventions to improve user engagement with online safety technology. 

\newpage
\bibliographystyle{apacite}
\bibliography{Safety_tech_citations}

\newpage
\appendix
\section{Appendix}

\subsection{Survey questions}
\label{sec:questions}

The full survey can be found on the OSF page for this project: \url{https://osf.io/zktny/}. Data supporting the study will be made available following publication. 

\subsection{Participant Demographics}
\label{sec:demographics}
Respondents were aged between 18 and 88, with a mean age of 44.64 (SD = 16.76). A total of 570 participants identified as female (53.4\%), 486 as male (45.54\%), and five as non-binary (0.468\%), with three selecting `prefer to self-describe' (0.28\%) and three selecting `prefer not to say' (0.28\%). The majority of respondents were White (910, 85.28\%). 478 respondents had degree level qualifications (44.79\%), 223 participants had non-degree level qualifications (vocational or similar) (20.89\%), 366 had no degree level qualifications (including completion of secondary school and below) (34.3\%).

\subsection{Descriptive Statistics}
\label{sec:app_desc_stats}

\begin{table}[!ht]
    \centering
    \begin{tabular}{lll}
        Variable & Mean & Standard Deviation \\ \hline
        Harm Concern & 68.12 & 23.4\\
        Political Orientation: More Right-wing (0-100) & 51.04 & 19.95 \\ 
        Digital Literacy & 77.82 & 14.67\\ 
        Age & 44.64 & 16.77 \\ 
        \\
         & Proportion Positive & \\ \hline
        Exposure: Witnessed Harmful Content & 0.67 & \\ 
        Exposure: Received Harmful Content & 0.31 & \\ 
        Gender: Female & 0.54 & \\ 
        Little Social Media Use & 0.39 & \\ 
        Education: No degree & 0.34 & \\ 
    \end{tabular}
    \caption{Decriptive statistics for independent variables}
    \label{tab:iv_desc}
\end{table}

\begin{table}[!ht]
    \centering
    \begin{tabular}{lllll}
        Technology Type & Technology & Aware? & N & Proportion \\ \hline
        Preventative (exposure) & Alter feed algorithm & No & 356 & 0.33 \\ 
        & Alter feed algorithm & Not sure & 130 & 0.12 \\ 
        & Alter feed algorithm & Yes & 581 & 0.54 \\ 
        & Deactivate feed algorithm & No & 622 & 0.58 \\ 
        & Deactivate feed algorithm & Not sure & 115 & 0.11 \\ 
        & Deactivate feed algorithm & Yes & 330 & 0.31 \\ \hline
        Preventative (posting) & Hide engagements & No & 499 & 0.47 \\ 
        & Hide engagements & Not sure & 94 & 0.09 \\
        & Hide engagements & Yes & 474 & 0.44 \\
        & Limit responses & No & 303 & 0.28 \\ 
        & Limit responses & Not sure & 77 & 0.07 \\ 
        & Limit responses & Yes & 687 & 0.64 \\ \hline
        Post-hoc & Block & No & 61 & 0.06 \\ 
        & Block & Not sure & 25 & 0.02 \\ 
        & Block & Yes & 981 & 0.92 \\ 
        & Report & No & 105 & 0.10 \\ 
        & Report & Not sure & 48 & 0.04 \\
        & Report & Yes & 914 & 0.86 \\ 
        & Unfollow & No & 73 & 0.07 \\ 
        & Unfollow & Not sure & 25 & 0.02 \\ 
        & Unfollow & Yes & 969 & 0.91 \\ 
    \end{tabular}
    \caption{Supporting Data for Figure 1, Proportion of People Aware of Each Technology}
    \label{tab:fig1_data}
\end{table}

\begin{table}[!ht]
    \centering
    \begin{tabular}{lllll}
        Technology Type & Technology & Aware? & N & Proportion\\ \hline
        Preventative (exposure) & Alter feed algorithm & Never & 100 & 0.17 \\ 
        & Alter feed algorithm & Not sure & 10 & 0.02 \\ 
        & Alter feed algorithm & Very rarely & 151 & 0.26 \\ 
        & Alter feed algorithm & Occasionally & 227 & 0.39 \\ 
        & Alter feed algorithm & Many times & 93 & 0.16 \\ 
        & Deactivate feed algorithm & Never & 57 & 0.17 \\ 
        & Deactivate feed algorithm & Not sure & 5 & 0.02 \\ 
        & Deactivate feed algorithm & Very rarely & 79 & 0.24 \\ 
        & Deactivate feed algorithm & Occasionally & 132 & 0.40 \\ 
        & Deactivate feed algorithm & Many times & 57 & 0.17 \\ \hline
        Preventative (posting) & Hide engagements & Never & 142 & 0.30 \\ 
        & Hide engagements & Not sure & 6 & 0.01 \\ 
        & Hide engagements & Very rarely & 128 & 0.27 \\ 
        & Hide engagements & Occasionally & 138 & 0.29 \\ 
        & Hide engagements & Many times & 60 & 0.13 \\ 
        & Limit responses & Never & 183 & 0.27 \\ 
        & Limit responses & Not sure & 8 & 0.01 \\ 
        & Limit responses & Very rarely & 146 & 0.21 \\ 
        & Limit responses & Occasionally & 200 & 0.29 \\ 
        & Limit responses & Many times & 150 & 0.22 \\ \hline
        Post-hoc & Block & Never & 166 & 0.17 \\ 
        & Block & Not sure & 5 & 0.01 \\ 
        & Block & Very rarely & 259 & 0.26 \\ 
        & Block & Occasionally & 338 & 0.34 \\ 
        & Block & Many times & 213 & 0.22 \\ 
        & Report & Never & 344 & 0.38 \\ 
        & Report & Not sure & 5 & 0.01 \\ 
        & Report & Very rarely & 235 & 0.26 \\ 
        & Report & Occasionally & 235 & 0.26 \\ 
        & Report & Many times & 95 & 0.10 \\ 
        & Unfollow & Never & 119 & 0.12 \\ 
        & Unfollow & Not sure & 7 & 0.01 \\ 
        & Unfollow & Very rarely & 209 & 0.22 \\ 
        & Unfollow & Occasionally & 379 & 0.39 \\ 
        & Unfollow & Many times & 255 & 0.26 \\ 
    \end{tabular}
    \caption{Supporting Data for Figure 2, Proportion of people who have used each technology (given awareness)}
    \label{tab:fig2_data}
\end{table}

\newpage
\clearpage
\subsection{Complete regression results}
\label{sec:app_reg_results}

\begin{table}[!htbp] \centering 
\begin{tabular}{@{\extracolsep{5pt}}lcccc} 
\\[-1.8ex]\hline 
\hline \\[-1.8ex] 
 & \multicolumn{4}{c}{\textit{Dependent variable:}} \\ 
\cline{2-5} 
 & Alter feed & Deactivate feed & Limit responses & Hide engagements \\ 
  & algorithm & algorithm & &  \\ 
\\[-1.8ex] & (1) & (2) & (3) & (4)\\ 
\hline \\[-1.8ex] 
 Harm Concern (scaled) & 1.110 & 1.312$^{*}$ & 1.050 & 1.394$^{**}$ \\ 
  & (0.091) & (0.123) & (0.080) & (0.109) \\ 
  & & & & \\ 
 Exposure: Witnessed  & 1.652$^{*}$ & 0.601 & 0.813 & 0.929 \\ 
 Harmful Content & (0.233) & (0.343) & (0.207) & (0.287) \\ 
  & & & & \\ 
 Exposure: Received & 2.000$^{***}$ & 1.307 & 1.306 & 2.128$^{***}$ \\ 
 Harmful Content & (0.203) & (0.262) & (0.184) & (0.220) \\ 
  & & & & \\ 
 Digital Literacy (scaled) & 1.114 & 1.004 & 0.949 & 0.879 \\ 
  & (0.089) & (0.119) & (0.079) & (0.101) \\ 
  & & & & \\ 
 Gender: Female & 0.842 & 1.109 & 1.429$^{*}$ & 0.940 \\ 
  & (0.181) & (0.232) & (0.160) & (0.208) \\ 
  & & & & \\ 
 Political Orientation: & 1.163 & 1.121 & 1.018 & 1.349$^{**}$ \\ 
 More Rightwing (scaled) & (0.090) & (0.118) & (0.079) & (0.102) \\ 
  & & & & \\ 
 Little Social Media Use & 0.652$^{*}$ & 0.668 & 0.881 & 0.404$^{***}$ \\ 
  & (0.206) & (0.282) & (0.180) & (0.270) \\ 
  & & & & \\ 
 Education: No degree & 0.752 & 0.751 & 1.047 & 0.911 \\ 
  & (0.192) & (0.253) & (0.170) & (0.219) \\ 
  & & & & \\ 
 Age (scaled) & 1.194 & 1.015 & 1.039 & 1.089 \\ 
  & (0.101) & (0.128) & (0.092) & (0.117) \\ 
  & & & & \\ 
 Constant & 0.881 & 2.101$^{*}$ & 0.930 & 0.701 \\ 
  & (0.253) & (0.371) & (0.229) & (0.309) \\ 
  & & & & \\ 
\hline \\[-1.8ex] 
Observations & 581 & 330 & 687 & 474 \\ 
Cox-Snell $R^2$ & 0.07 & 0.04 & 0.01 & 0.11 \\
\hline 
\hline \\[-1.8ex] 
\textit{Note:}  & \multicolumn{4}{r}{$^{*}$p$<$0.05; $^{**}$p$<$0.01; $^{***}$p$<$0.001} \\ 
\end{tabular} 
\caption{Supporting data for Figure 3 (preventative technologies)} 
  \label{tab:fig3_prevent} 
\end{table} 

\begin{table}[!htbp] \centering 
\begin{tabular}{@{\extracolsep{5pt}}lccc} 
\\[-1.8ex]\hline 
\hline \\[-1.8ex] 
 & \multicolumn{3}{c}{\textit{Dependent variable:}} \\ 
\cline{2-4} 
 & Unfollow & Block & Report \\ 
\\[-1.8ex] & (1) & (2) & (3)\\ 
\hline \\[-1.8ex] 
 Harm Concern (scaled) & 1.051 & 1.183$^{*}$ & 1.427$^{***}$ \\ 
  & (0.075) & (0.071) & (0.087) \\ 
  & & & \\ 
 Exposure: Witnessed Harmful Content & 1.629$^{**}$ & 1.781$^{***}$ & 2.027$^{**}$ \\ 
  & (0.176) & (0.172) & (0.215) \\ 
  & & & \\ 
 Exposure: Received Harmful Content & 1.885$^{**}$ & 2.009$^{***}$ & 2.437$^{***}$ \\ 
  & (0.194) & (0.175) & (0.174) \\ 
  & & & \\ 
 Digital Literacy (scaled) & 1.311$^{***}$ & 1.177$^{*}$ & 1.105 \\ 
  & (0.075) & (0.071) & (0.081) \\ 
  & & & \\ 
 Gender: Female & 1.404$^{*}$ & 1.580$^{**}$ & 1.382$^{*}$ \\ 
  & (0.153) & (0.145) & (0.162) \\ 
  & & & \\ 
 Political Orientation: More Rightwing (scaled) & 0.928 & 1.091 & 1.220$^{*}$ \\ 
  & (0.077) & (0.071) & (0.078) \\ 
  & & & \\ 
 Little Social Media Use & 0.591$^{**}$ & 0.542$^{***}$ & 0.384$^{***}$ \\ 
  & (0.160) & (0.153) & (0.183) \\ 
  & & & \\ 
 Education: No degree & 0.909 & 1.166 & 0.954 \\ 
  & (0.159) & (0.150) & (0.169) \\ 
  & & & \\ 
 Age (scaled) & 0.672$^{***}$ & 0.912 & 0.838 \\ 
  & (0.085) & (0.082) & (0.092) \\ 
  & & & \\ 
 Constant & 1.306 & 0.667$^{*}$ & 0.266$^{***}$ \\ 
  & (0.197) & (0.193) & (0.234) \\ 
  & & & \\ 
\hline \\[-1.8ex] 
Observations & 969 & 981 & 914 \\ 
Cox-Snell $R^2$ & 0.16 & 0.13 & 0.19 \\
\hline 
\hline \\[-1.8ex] 
\textit{Note:}  & \multicolumn{3}{r}{$^{*}$p$<$0.05; $^{**}$p$<$0.01; $^{***}$p$<$0.001} \\ 
\end{tabular} 
\caption{Supporting data for Figure 3 (post-hoc technologies)} 
  \label{tab:fig3_posthoc} 
\end{table} 
	
\end{document}